\documentstyle[preprint,aps,epsf]{revtex}

\def\be{\begin{equation}}
\def\ee{\end{equation}}
\def\ba{\begin{eqnarray}}
\def\ea{\end{eqnarray}}
\tighten
\begin{document}
\preprint{}
\draft
\title{Hadron-Gluon Interactions and the Pomeron}

\author{Eyal Gilboa}

\address{The University of Texas at
Austin, TX 78712-1081}

\date{\today}

\maketitle
\begin{abstract}
An interaction term is added to the QCD Lagrangian involving
hadron and gluon fields. An effective vertex is calculated for
such interactions through exchanges of reggeized gluons.  This
gives rise to an effective coupling for hadron-gluon elastic
scattering in the t-channel, which is used in an inclusive
hadron-hadron interaction from which the Pomeron intercept
$\alpha(0)_P$ is calculated.
\end{abstract}

\narrowtext

\section{Introduction}
 The QCD Pomeron and its Regge trajectory remains a formidable
 challenge in particle physics to this day.  Perturbative QCD has
 led to the BFKL equation \cite{Lipatov1} for which solutions have
 been constructed to obtain Regge behavior in particular the
 intercept of the Pomeron trajectory $\alpha(0)_P$. With this
 approach the scattering is done at the quark level requiring that a minimum of two gluons be exchanged to first
 order, and that all exchanges be done between the same two quarks with a color singlet projected from the $S$ matrix
 in each order.  One may refer to this treatment of a singlet exchange as a
`hard' Pomeron \cite{Ross} because the non-perturbative region of
QCD where $q<\Lambda_{QCD}$ is left out.
 In hadron physics this means that one assumes the separation of the scattering quarks to be much smaller than
 hadronic lengths.  Realistically, the dynamics of hadron scattering
is such that IR effects play a crucial role in predicting the
`soft' Pomeron \cite{Ross} behavior for which an observed
\cite{Landshoff} trajectory gives $\alpha_P(0)$ at approximately
1.08.  In a sense the Pomeron is an admission that perturbation
has its limits in describing the compositeness of hadrons from
constituent quarks, and it remains a challenge to formulate the
interplay between the two.  A more recent approach \cite{Ne'eman}
has emerged in which it has been shown that QCD in
 the IR region is a diffeomorphism like interaction in which two gluon
 fields when contracted with their color indices make up a tensor
 $G{\mu\nu}=\delta^{ab}A_{\mu}^aA_{\nu}^b$.  This tensor has properties similar to that of a metric tensor in
 gravity, which suggests new interaction terms to be included in Lagrangians
 involving both gluons and hadron matter fields.
 This approach assumes no information on the scattering
 mechanism of the quarks inside the hadron, but ensures that the hadron remains color neutral.

In the following, ideas from both approaches mentioned above will
be incorporated to treat the problem using perturbation (in which
case $q>\Lambda_{QCD}$) at the hadron level by introducing a term
involving two gluon fields with derivatives, coupled to hadron
fields. Although a BFKL equation will not be formulated at this
stage, we will consider reggeized gluon exchanges which will
ultimately lead to Regge behavior of the $S$ matrix.  By
introducing such interactions the model is an effective theory
which circumvents the problem of specifying which quarks
participate in the interactions, and thus there is no need to
assume that scattering is done only between the same two quarks.
The only stringent constraint is that the
color neutrality of the hadron is preserved in the scattering process.\\

Our starting point is with the following Lagrangian density:
\begin{equation}
{\mathcal L}={\mathcal L}_{QCD} + {\mathcal L}_{HG}\label{first}
\end{equation}
where the first term on the right denotes the full QCD Lagrangian,
while the second term denotes an interaction term given by:
\begin{equation}
{\mathcal L}_{HG}= \frac{\lambda}{2}
\bar{\phi}\delta^{ab}(\partial^{\mu}A^{\nu a}-\partial^{\nu}A^{\mu
a})(\partial_{\mu}A_{\nu}^{b}-\partial_{\nu}A_{\mu}^{b})\phi
\label{HG}
\end{equation}

\noindent In Eq.~(\ref{HG})\footnote{Greek letter indices will be
used to denote space-time, while small case Latin letter indices
will be used to denote color of the gauge group.  Also indices
pertaining to symmetries such as flavor have been suppressed.}
$\phi$ denotes a hadron matter field, $\lambda$ is a coupling with
dimension $M^{-D}$(where for example a scalar meson has $D=2$
while a spin $\frac{1}{2}$ baryon has $D=3$), $A^{\mu a}$ is a
gluon field and $\delta^{ab}$ is the $SU(N)$ color metric. Since
our ultimate goal is to investigate the Pomeron behavior, the
interaction term given above is the minimal interaction of gluons
coupled to a color neutral field which posses a Ward identity.
This will ensure current conservation, and the neutrality of the
hadron in the scattering process. Since the coupling $\lambda$ of
this interaction has a negative mass dimension, one should expect
that at high energy $M^D\lambda$ will decay.  In the language of
renormalization group \cite{peskin} ${\mathcal L}_{HG}$ consists
of irrelevant operators which die away at relatively high energy.
Since we are dealing with hadron interactions for which QCD is the
underlying theory then we should follow the energy scale defined
by this theory which is $\Lambda_{QCD}$. Thus for perturbation to
be useful we require that $\Lambda^D_{QCD}\lambda$ is small.

\section{Hadron-Gluon Vertex}
\label{sec2} 
\begin{figure}
\epsfxsize=3in \epsffile{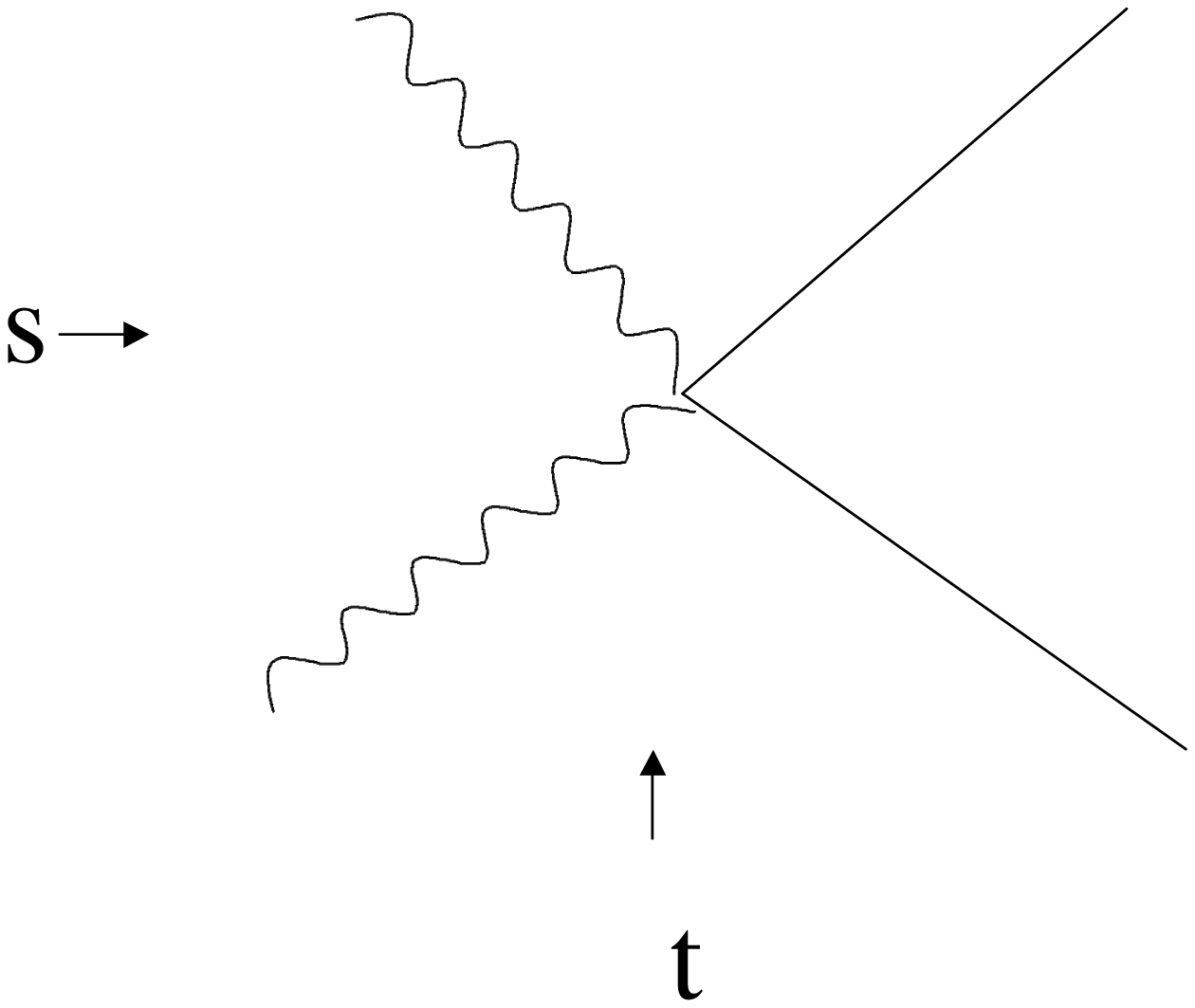} \caption{Tree level
amplitude for hadron-gluon scattering.} \label{dia1}
\end{figure}

The Feynman rule for the hadron-gluon interaction
follows from Eq.~(\ref{HG}):

\begin{equation}
V^{\mu\nu ab}=i\lambda \delta^{ab}\bigtriangleup^{\mu\nu}\\
\label{vertex}
\end{equation}
where
\begin{equation}
\bigtriangleup^{\mu\nu}=g^{\mu\nu}k \cdot k'-k^{\nu}k'^{\mu}
\end{equation}
$V^{\mu\nu ab}$ is the vertex shown in figure (\ref{dia1}).  For
simplicity the hadrons in (\ref{HG}) are chosen to be scalars,
though this could be extended to any hadron of any spin without
changing the tensorial structure of the vertex.  As was mentioned
above one can see that this vertex obeys a Ward identity;
$k_{\mu}V^{\mu\nu ab}=0$. This is crucial for ensuring the color
neutrality of the hadron.

\begin{figure}
\epsfxsize=3in \epsffile{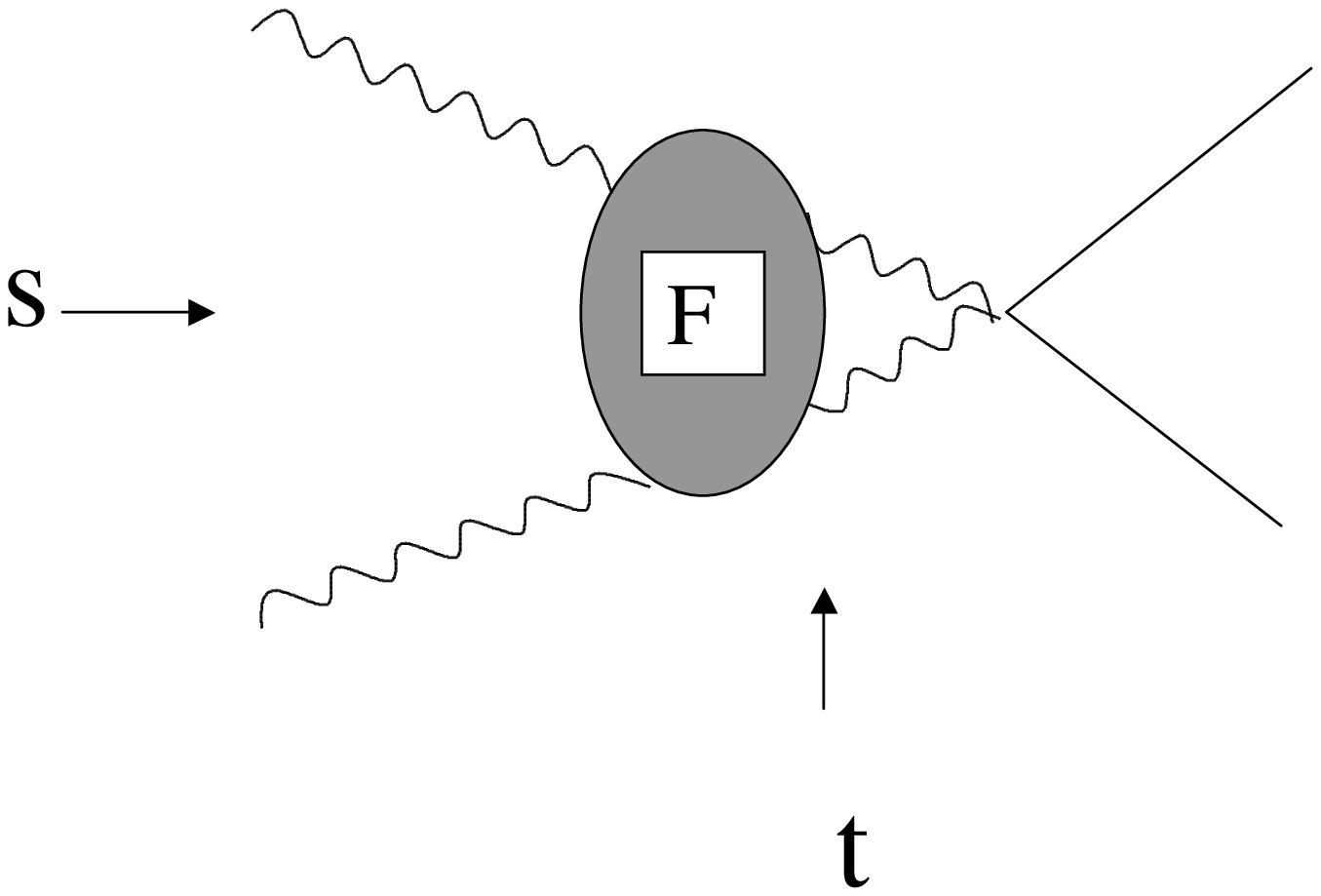} \caption{Multi-Regge
amplitude between two gluons.} \label{dia2}
\end{figure}

\begin{figure}
\epsfxsize=3in \epsffile{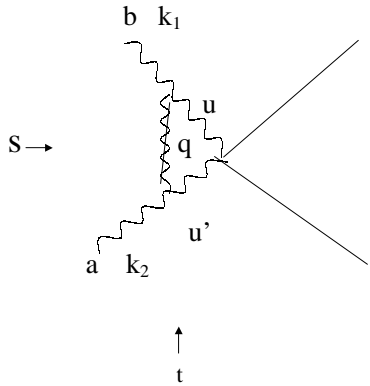}\caption{A reggeized gluon
exchange between gluons.}\label{dia3}
\end{figure}

Our next step is to evaluate corrections to this vertex in the
regge limit $(\frac{t}{s}\ll1)$ using perturbation theory where in
an s-channel process of two gluons annihilating to two hadrons, a
reggeized QCD octet exchange \cite{Lipatov} between the two
in-coming gluons takes place (figure~\ref{dia2}).\\
\noindent This octet exchange denoted by $F$ is given by
\cite{Ross}:
\begin{equation}
F=PT^a\otimes
T^a8\pi\alpha_s\frac{{\mathbf{k}}^2}{t}\left(\frac{s}{{\mathbf{k}}^2}\right)^{\alpha_G(t)}
\frac{1-e^{i\pi\alpha_G(t)}}{2}, \label{ra}
\end{equation}
where $P$ denotes the contracted on-shell gluon polarization
vectors, $T^a$ are the generators of the color group in the
adjoint representation, and $\alpha_G(t)$ is the regge trajectory
of the gluon. Expanding the second factor in powers of
$\alpha_G(t)$, the zeroth order reggeized amplitude is real, and
is given by an exchange of a reggeized gluon
\cite{Lipatov,Bartels,Gribov} in the t- channel between the two
incoming gluons (see figure \ref{dia3}, the vertical line
indicates that this a reggeized gluon). This amounts to replacing
the gluon propagator in the t-channel (Feynman gauge) with:
\begin{equation}
\frac{-ig^{\mu\nu}}{q^2} \Rightarrow
\frac{ig^{\mu\nu}}{\mathbf{q^2}}\left(\frac{s}{\mathbf{k}^2}\right)^{\epsilon{\left(\mathbf{q}\right)}},
\label{rgp}
\end{equation}
\newcommand{\T}{\left(\frac{s}{\mathbf{k}^2}\right)^{\epsilon{\left({\mathbf{q}}\right)}}}
\newcommand{\U}{\left(\frac{s}{\mathbf{k}^2}\right)^{\epsilon}}
\newcommand{\Ub}{\left(\frac{-s}{\mathbf{k}^2}\right)^{\epsilon}}
where
\begin{equation}
\epsilon\left(\mathbf{q}\right)=-\frac{N}{4\pi^2}\alpha_{s}\int\mathrm{d^2}\mathbf{k}\mathbf{\frac{q^2}{k^2(k-q)^2}},
\label{exp}
\end{equation}
and $\mathbf{k}$ is a typical momentum scale at which gluons
reggeize with the following condition:
\begin{eqnarray*}
\sqrt{s} \gg |{\mathbf{k}}|>\Lambda_{QCD}.
\end{eqnarray*}

We proceed to evaluate the first order amplitude using s-channel
unitarity which give rise to the Cutkosky cutting rules
\cite{Cutkosky}.  According to these rules the imaginary part of
an amplitude is given by:
\begin{equation}
{\rm{Im}}({\mathcal{A}}_{ab})=\frac{1}{2}(2\pi)^4
\delta\left(\sum_a p_a - \sum_b
p_b\right)\sum_{c}{\mathcal{A}}_{ac}{\mathcal{A}}_{cb}^{\dagger}
\end{equation}
where the summations in the brackets are over $in$ and $out$
states momenta respectively, and the third summation is over
intermediate states denoted by a cut in the diagrams where those
states appear. To obtain the real part of the amplitude one can
utilize the analytical properties of the S matrix \cite{Ross} in
which it can be shown that if in an s-channel cut the imaginary
part of the amplitude to $n^{th}$order is given by $A(\ln{s})^n$
in leading logs, then by using the identity
$ln{(-s)}=ln{(s)}-i\pi$ the real part of the amplitude is given by
$-\frac{A}{\pi(n+1)}(\ln{s})^{n+1}$.

\begin{figure}
\epsfxsize=3in \epsffile{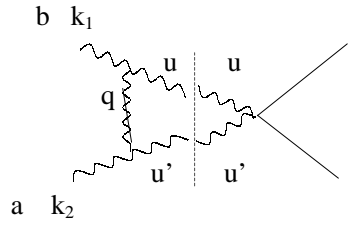}\caption{Imaginary part of
the first order amplitude for s-channel gluon-hadron scattering.}
\label{dia4}
\end{figure}

For the s -channel process at hand the zeroth order amplitude
(figure \ref{dia1}) is purely real.  The imaginary part of the
amplitude of the first order correction is given according the
cutting rules (figure \ref{dia4}):
\begin{eqnarray}
{\rm{Im}}({\mathcal{A}}_{HG})_1 & = &
\frac{1}{2}\sum_{G}\int\frac{\mathrm{d}^4
u}{(2\pi)^3}\frac{\mathrm{d}^4
u'}{(2\pi)^3}\delta(u^2)\delta(u'^2)(2\pi)^4\nonumber\\
 & \times & \delta^4(k+k'-u+u')({\mathcal{A}}_{GG})_0({\mathcal{A}}_{GH})_0^{\dagger} \label{foa}
\end{eqnarray}
where $({\mathcal{A}}_{GG})_0$ is the amplitude left of the cut,
and $({\mathcal{A}}_{GH})_0$ is the amplitude right of the cut,
and the sum is over intermediate gluon polarizations. In the limit
where the t-channel gluon with momentum $q$ is mostly transverse
and small compared to the center of mass energy, it is convenient
to introduce {\bf{Sudakov parameters}} and write $q^{\mu}$ as :
\begin{eqnarray}
q^{\mu}=\alpha k^{\mu}_1+\beta k^{\mu}_2+q_{\perp}
\end{eqnarray}
with $\alpha$, and $\beta$ being much less than unity.  Thus to a
good approximation it follows that:
\begin{eqnarray*}
q^2=-\mathbf{q^2}.
\end{eqnarray*}\\
With these simplifications the phase space of the integral in
Eq.~(\ref{foa}) becomes:
\begin{eqnarray}
\int\frac{\mathrm{d}^4 u}{(2\pi)^3}\frac{\mathrm{d}^4
u'}{(2\pi)^3}\delta(u^2)\delta(u'^2) (2\pi)^4\delta^4
(k+k'-u+u')\nonumber\\
 =\frac{s}{8\pi^2}\int
\mathrm{d}\alpha\mathrm{d}\beta\mathrm{d}^2\mathbf{q}\delta(-s\beta-\mathbf{q}^2)\delta(s\alpha-\mathbf{q}^2),
\label{ps}
\end{eqnarray}
where one delta function has been absorbed, and second order terms
in $\alpha$, and $\beta$ have been dropped.  The two remaining
delta functions in the integrand above give the condition that
$\alpha=-\beta=\frac{\mathbf{q}^2}{s}$.  Keeping in mind that $q$
is small the tree level amplitudes (figure~\ref{dia4}) is given
by:
\begin{eqnarray}
({\mathcal{A}}_{GG})_0 & = & 8\pi
s\alpha_{s}g^{\mu\lambda}g^{\nu\sigma}\varepsilon_{\mu}^
c\varepsilon_{\nu}^d\varepsilon_{\lambda}^b
\varepsilon_{\sigma}^e\nonumber\\
& \times & f^{abc}f^{ade}\left(\frac{1}{\mathbf{q^2}}\right)\T\\
({\mathcal{A}}_{HG})_0 & = &
\lambda\delta^{b'e'}\bigtriangleup^{\nu'\mu'}\varepsilon_{\nu}^{e'}\varepsilon_{\mu}^{b'},
\end{eqnarray}
with $f^{abc}$ being the structure constants of the color group.
Putting this together with Eq.~(\ref{ps}) into Eq.~(\ref{foa}),
and summing over gluon polarizations we get the following
expression for the imaginary part of the first order correction:
\begin{eqnarray}
{\rm{Im}}({\mathcal{A}}_{HG})_1 & = &
2N\delta^{be}\varepsilon_{\lambda}^
b \varepsilon_{\sigma}^e\bigtriangleup^{\lambda\sigma}\nonumber\\
& \times & \int\mathrm{d}^2{\mathbf{q}}
\frac{\lambda\alpha_{s}}{4\pi} \frac{1}{\mathbf{q^2}}\T.
\label{foa2}
\end{eqnarray}
\newcommand{\V}{\delta^{be}\epsilon_{\lambda}^{\gamma b
}(k)\epsilon_{\sigma}^{\delta
e}(k')\bigtriangleup^{\lambda\sigma}(k,k')}

\noindent The exponent in the integral of Eq.~(\ref{foa2}) is
given by Eq.~(\ref{exp}). The integral in the latter is infra-red
divergent, but can be regularized \cite{Ross} to give:
\begin{eqnarray}
\epsilon({\mathbf{q}})=-\frac{N}{2\pi}\left(\frac{1}{\varepsilon}-\gamma_E
+\ln{4\pi} +\ln{{\mathbf{q}}}\right)
\end{eqnarray}
The dependence of $\epsilon({\mathbf{q}})$ on regularization
parameters can be eliminated through scaling the expression above
by $\Lambda_{QCD}$.  This divergence appears at low energies where
gluons are confined, and QCD becomes non-perturbative.  This
region can be avoided by defining $\epsilon({\mathbf{q}})$ to
exist only where $q>\Lambda_{QCD}$. Thus, we redefine
$\epsilon({\mathbf{q}})$ as :
\begin{eqnarray}
\epsilon({\mathbf{q}})\rightarrow
\epsilon({\mathbf{q}})-\epsilon(\Lambda_{QCD})=-\frac{N}{2\pi}\alpha_s
\ln{\left(\frac{{\mathbf{q}}}{\Lambda_{QCD}}\right)}\label{epsilon}.
\end{eqnarray}

Including the running $\alpha_s$ \footnote{The running of
$\alpha_s$ can be included on the condition that the integration
over $\mathbf{q}$ in Eq.~(\ref{foa2}) be limited to the
perturbative region of QCD; namely from $\Lambda_{QCD}$ and up.
One cannot include the running in Eq.~(\ref{ra}) since there
$\mathbf{q}$ can take on any values including those not included
in perturbative QCD, and therefore would make $\alpha_s$ diverge
which would spoil the reggeization of the gluon.} in the
expression for $\epsilon({\mathbf{q}})$ Eq.~(\ref{epsilon})  we
get
\begin{equation}
\epsilon\left(\mathbf{q}\right) = -\frac{N}{b_o},
\end{equation}
 where $b_o$ is the familiar constant given by
\begin{eqnarray*}
b_o=11-\frac{2}{3}n_f,
\end{eqnarray*}
and $n_f$ is the number of flavors.  The $\mathbf{q}$ dependence
of $\epsilon$ has been eliminated, and Eq. (\ref{foa2}) reads:
\begin{eqnarray}
{\rm{Im}}({\mathcal{A}}_{HG})_1 & = & -\pi\epsilon
M(k_1,k_2)\left(\frac{s}{\mathbf{k^2}}\right)^{\epsilon}\int
\mathrm{d}{\mathbf{q}}\frac{2\lambda}{{\mathbf{q}}\ln\left({\frac{{\mathbf{q}}}{\Lambda_{QCD}}}\right)}\nonumber\\
& = &-\pi\epsilon M(k_1,k_2)\U\bar{\lambda}_{QCD}.\label{foa3}
\end{eqnarray}
$M(k_1,k_2)$ is the tensorial product of the vertex with gluon
polarization vectors times a color factor, and
$\bar{\lambda}_{QCD}$ is the corrected coupling of the original
vertex due to the running of $\alpha_s$.  Since we are in the
perturbative region of QCD away from $\Lambda_{QCD}$, the integral
above is well behaved as long as its lower limit does not include
this point.  Since we also anticipate that in the perturbative
region of QCD ~$\bar{\lambda}_{QCD}$ will be small compared to the
momenta of the gluons involved and fairly constant, we impose the
condition:
\begin{eqnarray}
\lambda=\bar{\lambda}_{QCD},
\label{norm}
\end{eqnarray}
where now $\lambda$ is assumed to be constant. This is just a
normalization condition on $\bar{\lambda}_{QCD}$ which says that
QCD corrections at high energies don't contribute much to the
running of $\lambda$ due to its negative mass dimension. According
to Eq.~(\ref{norm}) it follows that:
\begin{equation}
\int^{\sqrt{s}}_{\xi_o}\frac{\mathrm{d}{\mathbf{q}}}{{\mathbf{q}}\ln\left(\frac{{\mathbf{q}}}{\Lambda_{QCD}}\right)}
=\frac{1}{2},\label{defscale}
\end{equation}
where the lower limit is given by a new scale $\xi_o$ and the
upper limit is bounded by $\sqrt{s}$ since this is the maximum
momenta of the reggeized gluons.  The integral in
Eq.~(\ref{defscale}) can be integrated, and the value of $\xi_o$
can be obtained by the following:
\begin{equation}
\xi_o=\Lambda_{QCD}\left(\frac{\sqrt{s}}{\Lambda_{QCD}}\right)^\frac{1}{\sqrt{e}}.
\end{equation}
As expected $\xi_o$ is well above $\Lambda_{QCD}$, meaning we are
still in the perturbative region of QCD.
\\

With this the real part of the amplitude ${\mathcal{A}}_{HG}$ is
given by:
\begin{equation}
{\rm{Re}}({\mathcal{A}}_{HG})=M(k_1,k_2)\lambda\left(-1+\U\right).
\end{equation} Adding this to the zeroth term, the amplitude to
first order correction is given by
\begin{equation}
{\mathcal{A}}_{HG_{s-ch'
 0+1}}=M(k_1,k_2)\lambda\U(1-i\pi\epsilon)\label{parsch}.
\end{equation}
The second factor on the right of the equation above looks like it
contains the first two terms in a series of a multi-regge
exchange, suggesting that the entire amplitude in the s-channel
takes the form:
\begin{equation}
{\mathcal{A}}_{HG_{s-channel}}=M(k_1,k_2)\lambda\Ub.\label{finals}
\label{sch}
\end{equation}
To verify this we use crossing symmetry, and note that in the
t-channel where the process is the elastics scattering of a hadron
by a gluon (figure~\ref{dia3}) in which $t$ is negative, there are
no cuts to be made in a multi-regge exchange between the in-going
and out-going gluons, and the amplitude is purely real. Thus
according to Eq.~(\ref{parsch}) this amplitude up to first order
in $\epsilon$ is given by:
\begin{eqnarray}
{\mathcal{A}}_{HG_{t-ch' 0+1}} & = &
M(k_1,k_2)\lambda\left(\frac{-|t|}{{\mathbf{k}}^2}\right)^{\epsilon}(1-i\pi\epsilon)\\
& = &
M(k_1,k_2)\lambda\left(\frac{|t|}{{\mathbf{k}}^2}\right)^{\epsilon}e^{i\pi\epsilon}(1-i\pi\epsilon).
\end{eqnarray}
However, as mentioned before the t-channel amplitude is real to
all orders in $\epsilon$ suggesting that Eq.~(\ref{finals}) is the
entire multi-regge exchange amplitude for the process initially
considered.  In fact we may replace the vertex~(\ref{vertex}) with
the following:
\begin{equation}
V^{\mu\nu ab}=i\lambda
\delta^{ab}\bigtriangleup^{\mu\nu}\left(\frac{-Q^2}{{\mathbf{k}}^2}\right)^{\epsilon},
\label{newver}
\end{equation}
where $Q$ now is a measure of the gluons four momenta.  Such
corrections to the vertex are considered to be structure functions
of the hadron which become significant as the magnitude of $Q$
decreases due to $\epsilon$ being negative.

\section{Inclusive Interactions and the Pomeron}

Using the results obtained so far we now turn to treat the problem
of hadron-hadron scattering.  Our goal is to obtain the regge
intercept for such a process by calculating the amplitude to
second order in $\lambda$ together with the effective vertices
found in the previous section.

\begin{figure}
\epsfxsize=3in \epsffile{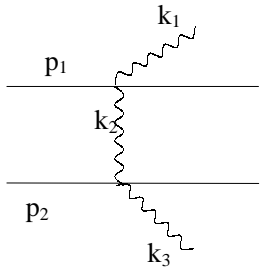}\caption{Hadron-Hadron
scattering with two out-going gluons.} \label{dia5}
\end{figure}

We consider an inclusive hadronic interaction by which two hadrons
with a center of mass energy $\sqrt{s}$ scatter through an
exchange of a gluon in the t-channel, plus two outgoing gluons
which ultimately give rise to jets~(figure \ref{dia5}).  Thus the
interaction is of the type:
\begin{eqnarray*}
h_1+h_2\rightarrow h'_1+h'_2+2g,
\end{eqnarray*}
where $h,h'$ are the initial and final hadronic states
respectively (these could include resonances), and $g$ denotes the
out-going gluons which would ultimately give rise to the emission
of two jets. Thus the process could very well describe the initial
stages of hadron diffraction dissociation in which the scattering
hadrons have a large rapidity gap.

\noindent Total momentum conservation of this process gives the
following:
\begin{equation}
p_1+p_2=p'_1+p'_2+ k_1 +k_3,\label{totmom}
\end{equation}
where $p_1,p_2$ and $p'_1,p'_2$ are the momenta of the in-coming
and out-going hadrons respectively, and $k_1,k_3$ are the momenta
of the out-going gluons. Conserving momenta at each vertex~(figure
\ref{dia5}) gives the following:
\begin{mathletters}
\label{vermom}
\begin{eqnarray}
p_1-k_2 & = & k_1+p'_2\\
p_2+k_2 & = & p'_2+k_3,
\end{eqnarray}
\end{mathletters}
where $k_2$ is the momentum of the gluon exchanged in the
t-channel which is assumed to be soft, and mostly transverse as
prescribed by the regge limit.  Therefore we may assume that its
$t$ and $z$-components are negligible, so that to a good
approximation we have:

\begin{mathletters}
\label{conb}
\begin{eqnarray}
k^{\mu} & \rightarrow & k^{\mu}_{\perp}\\
\frac{{\mathbf{k}}^2_2}{s}& \ll & 1.
\end{eqnarray}
\end{mathletters}

Working in the center of mass frame (assuming again that our
hadrons are scalars) in the massless limit where we have~
$p_1=(\frac{\sqrt{s}}{2},\frac{\sqrt{s}}{2},0,0),
p_2=(\frac{\sqrt{s}}{2},-\frac{\sqrt{s}}{2},0,0)$~ it is again
convenient to introduce the {\bf{Sudakov parameters}} and write
the following vectors as:
\begin{eqnarray}
\label{sudamom} k_i^{\mu} &=&
\alpha_ip_1^{\mu}+\beta_ip_2^{\mu}+k_{i\perp}^{\mu},
\end{eqnarray}
where $\alpha_2=\beta_2=0$.
\\
\noindent In specifying the kinematic regime for the process at
hand, we note that since the reggeon exchanged (in this case it is
a gluon field) between the two hadrons is soft relative to the
in-coming energy, the process is plagued with non-perturbative
effects (the exchange is done at long distances). This requires
the knowledge of parton distribution functions in this region for
which at this time we don't have a way of computing from first
principles. Never the less, for perturbative purposes we will
assume that $k_2\gg m_h \approx \Lambda_{QCD}$ which ensures that
\begin{eqnarray*}
\frac{1}{k_2}\approx\Delta t_{int}\ll \Delta
t_{hadron}\approx\frac{1}{m_{hadron}}.
\end{eqnarray*}
In doing so we adopt the scheme \cite{Gribov} that a hadron is a
composite of fields that have a life time much greater than their
interaction time; such that the interaction described
(figure~\ref{dia5}) is just the first step in hadron diffraction
dissociation where an off-shell gluon field probes the hadron in a
way that initially leaves it in tact, though eventually will lead
to it's break up. Since gluons emitted in the first stages of the
brake up will eventually hadronize, there is no real way to
distinguish them from the out-going hadrons in terms of their
momenta after the break up has occurred. Since an inclusion of
these gluons is a must in order to keep the color neutrality of
the hadron, we could very well consider them as a partons carrying
some fraction of the in-coming energy.  We assume this fraction is
small such that:
\begin{mathletters}
\label{conc}
\begin{eqnarray}
1\gg\alpha_1>\alpha_3 \gg \frac{{\mathbf{k}}_2^2}{s}\\
1\gg\beta_3>\beta_1\gg\frac{{\mathbf{k}}_2^2}{s}.
\end{eqnarray}
\end{mathletters}
The orderings above conform to the requirement that initially the
hadrons momenta are not altered much (in these first stages the
hadrons are still intact), meaning there is a high rapidity gap in
this stage, which enforces the strong ordering on the left of
(\ref{conc}). Now since the out-going gluons are on shell we have
the condition $s\alpha_1\beta_1={\mathbf{k}}_1^2$ (with a similar
one for the second on-shell gluon with momentum $k_3$). Due to
this and (\ref{vermom}) we see that $k_{1\perp},k_{3\perp}\approx
k_{2\perp}$, and thus at most
$\alpha_i\beta_i\approx\frac{{\mathbf{k}}_i}{\sqrt{s}}$ which
justifies the strong ordering on the right of (\ref{conc}).

Having specified the kinematics, Eq.~(\ref{newver}) can be used to
evaluate the zeroth order amplitude of hadron-hadron scattering
with two out-going gluons to give:
\begin{eqnarray}
{\mathcal{A}}_{(HHG)} & = &
\delta^{ab}\lambda\lambda'\bigtriangleup^{\mu\lambda}
\left(\frac{-q_1^2}{{\mathbf{k}}^2}\right)^{\epsilon}
\left(\frac{g_{\lambda\sigma}}{k_1^2}\right)\nonumber\\
& \times &
\bigtriangleup^{\sigma\nu}\left(\frac{-q_2^2}{{\mathbf{k}}^2}\right)^{\epsilon}
\varepsilon_{\mu}^{a}\varepsilon_{\nu}^{b}
 \label{fa}
\end{eqnarray}
Since this is a tree level amplitude which is real meaning there
are no cuts to be made at all on the diagram, we have the momenta
$q^2_1=-(k_2+k_1)^2$, and $q^2_2=-(k_2-k_3)^2$ that are obtained
from each vertex in figure~(\ref{dia5}) respectively. Using the
on-shell condition for both $p'_1$, and $p'_2$ in the massless
limit we get:
\begin{eqnarray}
(-q^2_1)^{\epsilon}=s^{\epsilon}\beta_1^{\epsilon}\nonumber\\
(-q^2_2)^{\epsilon}=s^{\epsilon}\alpha_3^{\epsilon}.
\label{finalver}
\end{eqnarray}
The tensorial product in Eq.~(\ref{fa}) is
given by:
\begin{eqnarray}
g_{\lambda\sigma}\bigtriangleup^{\mu\lambda}\bigtriangleup^{\sigma\nu}
 & = & \Big(g^{\mu\nu}(k_1\cdot
k_2)(k_2\cdot k_3)\nonumber\\
 & - & (k_1\cdot
k_2)k_2^{\nu}k_3^{\mu}-(k_2\cdot k_3)k_2^{\mu}k_1^{\nu} +
(k_1\cdot k_3)k_2^{\mu}k_2^{\nu}\Big). \label{tensor}
\end{eqnarray}
Again by use of (\ref{conb}), and (\ref{conc}) in the massless
limit, keeping only second order terms in Sudakov parameters the
expression above is reduced to:
\begin{eqnarray}
g_{\lambda\sigma}\bigtriangleup^{\mu\lambda}\bigtriangleup^{\sigma\nu}&
= & -\frac{g^{\mu\nu}}{4}\alpha_3\beta_1s^2
\end{eqnarray}
From the discussion at the end of section \ref{sec2} the
correction to the hadron-gluon vertex given by the factor in
Eq.~(\ref{newver}) can be associated with hadron structure
functions due to reggeized gluon exchanges in the perturbative
region of QCD.  In light of this and Eq.~(\ref{finalver}) it is
useful to define the normalized longitudinal gluon distribution
functions in terms the Sudakov parameters given by:
\begin{eqnarray}
f_i=\psi_i^{\dagger}(\alpha_i)\psi_i(\alpha_i)=(1+\epsilon)\alpha_i^{\epsilon},
\label{distfunc}
\end{eqnarray}
where $\psi(\alpha_i)$ is the amplitude of a gluon with a
fractional longitudinal momentum $\alpha_i$.  Thus the probability
of finding a gluon with a longitudinal momentum between
$\alpha_i$, and $d\alpha_i$ is given by
\begin{eqnarray*}
dP=f_id\alpha_i.
\end{eqnarray*}

\noindent Using (\ref{finalver}), (\ref{tensor}), and
(\ref{distfunc}) the amplitude (\ref{fa}) may be written as:
\begin{equation}
{\mathcal{A}}_{HHG}=\varepsilon^a_{\mu}\varepsilon^b_{\nu}<\alpha_3><\beta_1>
\frac{g^2g^{\mu\nu}}{{\mathbf{k}}_2^2}\left(\frac{s}{{\mathbf{k}}^2}\right)^{2+2\epsilon},
\label{faa}
\end{equation}
where $<\alpha_3>,<\beta_1>$ are the expectation values of the
gluons longitudinal fractional momenta $k_3$ and $k_1$
respectively, and $g^2$ is a dimensionless coupling given by
$g^2=\frac{\lambda\lambda'{\mathbf{k}}^4}{4(1+\epsilon)^2}$.  Thus
the amplitude (\ref{faa}) is factored into two parts; the first
dealing purely with the out-going gluon momenta and their
polarizations, while the second is a reggeized propagator which
describes an exchange of a reggeon between two hadrons with a
regge trajectory given by:
\begin{equation}
\alpha(t)=2+2\epsilon. \label{tra}
\end{equation}
This trajectory is independent of $t$ though this should be
expected since we have considered only the tree level amplitude in
our scheme.  We expect that in a completely elastic process where
the amplitude (\ref{fa}) is a cut of an imaginary amplitude of the
type $h_1+h_2\rightarrow h'_1+h'_2$ with both s,t-channel gluons
appearing in the intermediate states, higher order corrections
will lead to a trajectory with $t$ dependence. Since the amplitude
(\ref{faa}) is color neutral meaning our S-matrix describes
transitions between $in$ and $out$ states with no net change in
color nor in flavor, we conclude that the trajectory in
(\ref{tra}) is nothing more than $\alpha_P(0)$ which is the regge
intercept of the Pomeron. It is well known from regge theory
\cite{Collins} that at high energy for a particular reggeon
exchange it is $\alpha_R(0)$ which dominates the interaction, and
gives the asymptotic behavior of the total cross section namely:
\begin{eqnarray*}
\sigma_{total}\sim s^{\alpha_R(0)-1}.
\end{eqnarray*}
For $\alpha_R(0)>1$ the Pomeranchuk theorem \cite{Pomeranchuk}
predicts regge exchanges of particles carrying quantum numbers of
the vacuum (gluons), and a rising cross section.  Eq.~(\ref{tra})
can be written as:
\begin{equation}
\alpha(0)_P=2+2\epsilon=2-\frac{2N}{11-\frac{2}{3}n_f}.
\label{int}
\end{equation}
Putting $N=3$, and $n_f=6$ in Eq.~(\ref{int}) we get in
$\alpha_P(0)=1.14$, giving a rising cross section in accord with
the Pomeranchuk theorem. A peculiar feature of this expression is
that it does not depend on the strong coupling $\alpha_s$.  We
stress that this comes about because the scale chosen in
Eq.~(\ref{epsilon}) was precisely the scale at which
non-perturbative effects become important namely $\Lambda_{QCD}$.
This gave a cancellation of the coupling up to a constant. Had a
different scale $M$ been chosen which coincides with a particular
hadron mass, such a cancellation would not have occurred, and an
explicit dependence on the coupling would have appeared in the
expression for $\alpha_P(0)$. For light hadrons the sale
$\Lambda_{QCD}$ is a good approximation for describing their mass
scale, and in doing so we have simplified the calculation of
$\alpha_P(0)$.

\section{Conclusions}
In the calculation for $\alpha_P(0)$ above it seems that a
significant contributing factor for it's value comes from knowing
the corrected vertices Eq.~(\ref{newver}) which provide knowledge
on hadron structure functions in the regge region. It is important
to stress that these functions are in the region where
$q>\Lambda_{QCD}$ and that complete knowledge of hadron structure
functions in the non-perturbative region are still not at hand. In
our model a `soft' Pomeron is an exchange of a single gluon in the
t-channel with two out-going gluons forming a color singlet in
their final state.  The term `soft' is used because $q$ is low
(unlike the `hard' Pomeron in which scattering is treated at the
parton level between two partons requiring high $q$), and also
because this model treats the hadron as an effective field in
which color singlet exchanges are more likely to happen between
different partons since the exchange is a soft one (long
distances), and the non locality of the hadron is taken into
account through it's structure functions. The measured value of
$\alpha_P(0)$ is 1.08 \cite{Landshoff}, so we conclude that the
structure functions in the non-perturbative region play a crucial
role in understanding the `soft' Pomeron intercept.

The dependence of $\alpha_P(0)$ (Eq. (\ref{int})) on the number of
flavors has an interesting implication. As the number of flavors
increases $\alpha_P(0)$ drops.  In fact increasing the number of
flavors to seven would cause it to drop below 1.  Having
$\alpha<1$ would clash with Pomeranchuk's theorem which guarantees
a rising cross with $s$ as long as the exchanges are of particles
which carry quantum numbers of the vacuum. In other words if we
are to uphold the theorem even at very high energies, six flavors
seems to be a crossroads in our model, since with the addition of
one more flavor some new physics should appear to compensate for
the drop in $\alpha$. Whatever this new physics is
(Supersymmetry?) it should involve particles that reggeize, and
that play a major role in hadron interactions.

Finally the tensor given in (\ref{HG}) of gluons and their
derivatives before contraction on its Lorenz indices has the form
$R_{\mu\nu\rho\sigma}$.  This tensor is antisymmetric on the
exchange of $\mu$ and $\nu$, or $\rho$ and $\sigma$, but is
symmetric with the exchange of $\mu\nu$ with $\rho\sigma$.  These
algebraic properties of a tensor describe a particle of spin 2
\cite{Weinberg}.  Eq. (\ref{int}) would have given $\alpha=2$ had
we not incorporated loop corrections in our amplitude.  This
conforms to the model given in \cite{Ne'eman} which predicts a
spin 2 behavior in QCD with no matter fields.

\section{Acknowledgement}
I would like to thank professor Yuval Ne'eman for his helpful
comments.


\end{document}